\documentclass[num-refs]{wiley-article}
\usepackage{siunitx}

\papertype{Research Article}
\title{Nonparametric optimization of short-term confidence bands for wind power generation}% in Uruguay}

\abbrevs{PLF, Plant Load Factor. }%DEF, doesn't ever fret; GHI, goes home immediately.}

\author[1\authfn{1}]{Claudio RISSO, PhD}
\author[2]{Gustavo GUERBEROFF, PhD}
\contrib[\authfn{1}]{Corresponding author}

\affil[1]{Computer Science Institute, InCo, UdelaR, Montevideo, P.C. 11300, Uruguay, crisso@fing.edu.uy}
\affil[2]{Probability and Statistics Laboratory, IMERL, UdelaR, Montevideo, P.C. 11300, Uruguay, gguerber@fing.edu.uy}

\corraddress{Claudio Risso, InCo, Facultad de Ingenier\'{\i}a, UdelaR, Montevideo, P.C. 11300, Uruguay}
\corremail{crisso@fing.edu.uy}

\fundinginfo{ANII (Agencia Nacional de Investigaci\'on e Innovaci\'on), llamado Fondo Sectorial de Energ\'{\i}a (FSE) 2015, ANII-FSE\_110454}% Funder Two, Funder Two Department, Grant/Award Number: 123459}

\runningauthor{C. Risso et al.}

\newcommand\vs{\vspace{0.05in}}
\newcommand\secref[1]{Section \ref{#1}}
\newcommand\figref[1]{Figure \ref{#1}}
\newcommand\tabref[1]{Table \ref{#1}}
\newcommand{\headcell}[1]{\cellcolor{black!20}{\bf #1}}

\begin{document}

\maketitle

\begin{abstract}
\small{The increasing rate of penetration of non-conventional renewable energies
is affecting the traditional assumption of controllability over energy sources.
Power dispatching scheduling methods need to integrate the intrinsic randomness of some new sources,
among which, wind energy is particularly difficult to treat.
This work aims on the construction of confidence bands around wind energy forecasts.
Complementarily, the proposed technique can be extended to integrate multiple forecasts into a single one,
whose band's width is narrower at the same level of confidence.
The work is based upon a real-world application case,
developed for the Uruguayan Electricity Market,
a world leader country in the penetration of renewable energies.}
% Please include a maximum of seven keywords
\keywords{wind power, non-conventional renewable energy, forecasting, confidence bands, combinatorial optimization}
\end{abstract}

%%%%%%%%%%%%%%%
%%% INTRODUCTION %%%
%%%%%%%%%%%%%%%
%%%%%%%%%%%%%%%
%%% INTRODUCTION %%%
%%%%%%%%%%%%%%%
\section{Introduction}

Whether due to economic pressure or environmental concerns,
the rate of penetration of non-conventional renewable energies has been increasing rapidly over recent years,
and it is expected to grow even faster in the years to come.
Short-term operation and maintenance of electrical systems
relies on optimal power dispatch scheduling methods.
Either renewable or not, conventional energy sources are dispatchable on request, i.e.,
authorities can control when and how much power will be provided from each source.
Conversely, non-conventional renewable energies are not controllable, are intermittent and uncertain,
even within a few hours period ahead.
The intrinsic stochastic nature of the new energy sources
turns out the short-term dispatch of the grid into a much harder challenge,
which necessarily must coexist with randomness coming from significant portions of the installed power plant.
This work regards with the crafting of confidence bands for the wind power generation.
It is based on an application case of Uruguay, a worldwide leader in the usage of renewable energies,
but obviously it can be ported to another system or country.\vs

In the context of forecast of daily scenarios for time series
there exist a vast literature and plenty of methods proposed to accurately predict a likely behavior for the stochastic process involved.
Many of them use purely statistics techniques, both parametric and nonparametric, in order to infer forecasts.
As an example of a novel nonparametric approach to infer forecasts we mention the Pattern Sequence-based Forecasting algorithm (PSF) \cite{5620917}.
This method has provided promising results when applied to electricity prices and electricity demand time series forecasting in several international markets,
at a horizon of one or a few days ahead. The main idea of the PSF algorithm and more recent variants involves three parts working sequentially:
1) the historical of time series is clustered in groups that have similar temporal daily profiles;
2) the time series of daily profiles is converted into a discrete sequence that labels the clusters according on which day they belongs to;
3) given the label corresponding to the current day,
a window of days of fixed length is used to search along the whole past sequence of labels to match those with the same sequence as the window ending at present time:
the profile for the day to be predicted is constructed averaging the profiles that follows to each one of those identical windows in the past. Analogously there exist several parametric and semi-parametric models, most of them using standard and well-known techniques coming from the analysis of time series for the daily forecasting electricity prices and electricity demand (see for example \cite{WERON2008744}).
Statistical methods like those perform well to forecast: electricity prices, electrical demand time series,
hydraulic contributions to water reservoirs, or even to forecast mid and long-term availability of wind power.
When the application regards with short-term wind power generation, however,
the accuracy of such statistical methods degrades notably,
even over short periods of time that range from a few hours till two or three days ahead. The reason is that temporal variations in the series of wind power are strongly more fluctuating than electricity prices and electrical demand time series (and the other examples mentioned above).\vs

%Of course, there also exists a large arsenal of sophisticated mixed models involving exogenous variables.
%But, as we have commented before,
%the variability and irregularity of wind power time series prevent us to use these methodologies because the errors obtained with our data set are out of our goal. \vs
%
%The methodology that we have constructed, and that is the main contribution of this paper, uses as imput a cluster analysis for the daily patterns of historical wind power in the region. As was expected we found four profiles for the clusters with a comparable number of members, wich are interpreted in a simple way: 1) very windy days; 2) windless (or almost windless) days; 3) windless in the first half of the day and windy in the second half; 4) windy in the first half of the day and windless in the second half. In this way we can consider a kind of markovian process whose states are the daily clusters.
%As we will explain in detail in the following sections, this cluster separation allows us to construct bands of errors of better quality.
%A smarter approach incorporates information related to the historical deviations of the actual process with respect to the forecast.
%Several works have been done in this address (REFERENCIAS), including parametric and nonparametric techniques.

Complementarily, there are approaches for short-term wind power forecasting based on numerical simulations of atmosphere's wind flows.
For a day ahead period, or even larger time windows,
numerical simulations are usually more accurate than purely statistical models.
Such models are deterministic, while the underlying physical phenomena is chaotic by nature.
So, they perform better than purely statistical methods to follow the process whereabouts at early stages,
but are far from being trustworthy in what respects to the construction of likely scenarios at larger times.
Summarizing: the scheduling of short-term dispatch must coexist with randomness, so,
even though wind's simulations provide valuable information,
they must be enriched to account the inherent stochasticity of the process.\vs

Regarding the wind power predictions, our article neither explores the application of purely statistics methods,
nor the outcome of deterministic simulations.
Rather, it is based in the hybridization of both strategies (deterministic forecasts provided as input combined with a statistical optimization analysis to construct confidence bands) to get a more comprehensive one.
The idea, then, is to design a statistical model based on the deterministic forecasts as a mainstream of what to expect, and incorporate information related to the historical deviations of the actual process with respect to the corresponding forecast. Assuming persistent behavior, we can estimate such deviations based on how they were over a previous training period or set.\vs

Several recent papers have been issued in this same line of work. In \cite{FOR:FOR2367} parametric processes guided by stochastic differential equations are introduced. These processes involve a drift term,
which acts as a force that tends to attract the process towards the forecast (which is known as an input),
and also involve a diffusive term with a Wiener process factor as usual.
Different parametric models are studied by considering specific forms for the drift and diffusion terms. 
The basic idea of the mentioned work consists in approximate these parametric non-Gaussian stochastic processes by corresponding Gaussian processes with the same mean,
variance and covariance structure. Such approximations allow the estimation of the parameters through maximum likelihood techniques.
%including parametric and nonparametric techniques.
Following the ideas of these authors, and using the same data set for wind power in Uruguay as in this work,
%in \cite{EKTempone} confidence bands are built where the current process should belong with a highly reasonable probability.
\cite{EKTempone} synthesizes multiple realizations of the calibrated process in order to build confidence bands.
However, the differences observed between actual data and computed bands from the forecasts, and the width of the bands,
exceeds the necessary standards for the practical applications that have motivated our work.\vs

In \cite{WE:WE2129} methods that also use deterministic forecasts as inputs are introduced, and through a subtle analysis that involves the historical data to estimate nonparametric forecast error density for some relevant times chosen appropriately, the authors have succeeded in generating wind power scenarios with their respective probabilities. \vs

Conversely to \cite{WE:WE2129} goals, our interest is not focused on particular trajectories and their probabilities but aims on crafting confidence bands around wind power forecasts in such a way that the energy outside the bands (in a given metric specified in \secref{sec:bandopt}) is precisely bounded.
At this respect, our article presents a novel approach. The method is purely non-parametric, since it makes no assumptions upon the physical phenomena,
nor any hypothesis about the involved random processes in the time series (conditions on homogeneity in time, seasonal behavior of the temporal series or any kind of markovian hypotheses, are not necessary for this work).
It is based upon a mixed integer optimization problem,
which aims on getting the narrowest band around a forecast
that keeps the off-band aggregated energy below a given threshold.
The trustworthiness of this construction relies on the calibration of bands over a historical training set.
As an innovation, to allow the optimization to go as farther as possible,
the model enables to discard up to a given percentage of training samples
which are treated as \emph{atypical profiles}.
At first glance this last feature might look risky, in the sense that, in advance,
one cannot tell whether the day to come will match or not an atypical profile.
However, as we will see later,
when the method is trained with forecasts coming from more than one provider,
whenever they are conditionally independent or weakly dependent,
using a combination of them and their bands results in a more accurate outcome than each construction computed by separate.\vs

The structure of this article matches the stages of the novel technique.
\secref{sec:bandopt} presents the optimization model to create likely bands of minimum width as well as results from experimental evaluation,
while \secref{sec:combfore} shows how a combination of bands calibrated up from independent forecasts performs better than any of them by separate.
Finally, \secref{sec:concl} summarizes the main results and future lines of work.

%Usually fluctuations in the actual data are large and can be noticeably appart from the forecast, making of studies based only in deterministic forecast are not satisfactory. This is the main reason to include an analysis of the errors to improve on the performance of the predictions and the design of ditribution of electrical power for the short-term future.
%

%%%%%%%%%%%%%%%%%%%%%%%%%%%%%
%%% OPTIMIZATION OF CONFIDENCE BANDS %%%
%%%%%%%%%%%%%%%%%%%%%%%%%%%%%
%%%%%%%%%%%%%%%%%%%%%%%%%%%%%
%%% OPTIMIZATION OF CONFIDENCE BANDS %%%
%%%%%%%%%%%%%%%%%%%%%%%%%%%%%
\section{OPTIMIZATION OF CONFIDENCE BANDS}\label{sec:bandopt}

Optimization of confidence bands is in the core of this framework.
We provide an expression to compute a band around any forecast
and, for that concrete formula,
we seek for the narrower band that satisfies a set of constraints,
which impose limits to the actual process in its deviations
in accordance with the historical behavior.\vs

The information required to determine an instance comprises the following data sets.
In the first place, we need a historical of wind power forecasts.
We consider a collection ${\mathcal P}$ of deterministc registers that involves short-term point forecasts over a horizon of a few days ahead;
i.e., a family of vectors $p^d \in [0,1]^T$, with fixed $T$, which is set by the number of samples along the time horizon.
Here, $d$ is the index for each day on which the construction of the bands begin; $d \in D$, the set of daily historical observations. 
Wind power forecasts usually span from one up to three days,
i.e., from 24 to 72 hours, and time is discretized at a rate of one sample per hour.
Let $T-1$ be the limit of hours ahead available for each forecast.
We assume that all forecasts share the same time horizon,
and that $t=0$, the current power, is the only data known for sure.
For simplicity, the wind power is expressed as the Plant Load Factor (PLF),
which corresponds to the actual power generated at each time,
divided by the sum of the installed power capacity of each wind turbine in the system at each moment.
Hence, information is normalized so we can disregard changes in the installed capacity during the period of analysis.
Thus $p_t^d\in [0,1]$ is the normalized point forecast of the wind power $t$ hours ahead,
within the vector associated to the forecast issued on the day $d$.\vs

The second part of the input data set comprises the actual historical wind power time series samples,
grouped into a collection, ${\mathcal{W}}$,
whose elements $w^d \in [0,1]^T$ are also assumed normalized.
Hence, $w_t^d\in [0,1]$ is the actual PLF measured $t$ hours after the beginning of the day $d$.
For consistency, since the current state can be measured rather than forecasted,
$p_0^d=w_0^d$ for each day $d$.
Observe that the set ${\mathcal{W}}$ usually has duplicated records,
for instance: $w_{24}^d=w_0^{d+1}$.
Despite that, we have chosen this format to simplify those expressions that link with forecast information.
Regarding forecasts, however, the previous equality doesn't hold.
In fact, $p_{24}^d$ (a sample, forecasted 24 hours ahead) is different from $p_0^{d+1}$ (the actual value measured a day later).\vs

It is clear that, given any two bands containing the real process inside of them at the same instants, the narrower band is of better quality.
Wind power generation is a process hard to anticipate,
and violations to computed bands is a fact we must coexist with.
However, not every violation has the same severity
in terms of its impact to the power grid.
In the context of the short-term energy dispatch,
how much cumulated energy falls down outside the band
is a convenient metric to asses the confidence of the pair: forecast plus computed band.
In this work we use the following expression as a metric for the reliability\footnote{The expression on the right hand side corresponds precisely to the anti-reliability,
which of course is the complement of the reliability; hence the notation for the left hand side.} ${\mathcal R}^{d}$ of a band around a given forecast $p^{d}$:
$$1 - {\mathcal R}^{d}(w,lb,ub)=\frac{1}{T}\sum_{t=0}^{T-1} \left(\max \left[w(t)-ub\left(p^{d},t\right),0\right]+\max \left[lb\left(p^{d},t\right)-w(t),0\right] \right),$$
where $ub$ and $lb$ respectively are the functions that determine upper and lower limits for the bands along the forecasted period,
and $w$ is the actual generation, known once reality is revealed.
Functions $lb$ and $ub$ take a forecast ($p^{d}$) and an instant ($t$) as their inputs,
while their outputs are the respective bounds to expect. \vs
%In other words, $1-{\mathcal R}_T \in [0,1]$ represents the tolerance value %for the mean energy (over the $T$ points which are the domain of forecast %%$p^d \in {\mathcal P}$) which is allowed for the violation of the bands, as %explained in detail below. \vs

As mentioned, the feasible region of the optimization model imposes limits to the severity of violations to the confidence band.
Besides, in order to improve the quality as further as possible,
the model allows to discard up to a limit of elements in the training set,
which are atypical, specially bad forecasts that whether included would either:
deteriorate the accuracy of the result, or force us to use too broad bands.
So, to complete an instance we must set values to those quantities.
The parameter $\theta \in [0,1]$ limits the amount of energy allowed to fall down outside the band along the optimization horizon.
The parameter $\lambda \in [0,1]$ sets a minimum fraction of \emph{regular} (i.e. not atypical) forecasts to be used in the effective training set
or, in other terms, $(1-\lambda)$ is the maximum fraction of atypical days allowed to be discarded.
It is worth mentioning that the limit for off-band energy only accounts over regular forecasts.

%%% Minimal relative bandwidth version %%%
\subsection{Minimal relative width of bands}

This work considers those confidence bands defined by relative deviations with respect to forecasted values, which are simple to calculate and optimize, and yet lead to accurate results.\vs

Let $\{x_t\ge 0\}$ be a set of coefficients associated to the time series analyzed,
which delimits the width of the band. That is, for any instant $t$ within the time horizon of the forecast issued on day $d$,
we take $p^d_t$ and compute the lower and upper limits of the band using the expressions
$\max \left[0,(1-x_t)p^d_t\right]$ and $\min\left[1,(1+x_t)p^d_t\right]$ respectively.
Hence, $\{x_t: 0\le t\le T-1\}$ comprises the first set of control variables that modulates the relative width of the band for a given forecast $p^{d} \in {\mathcal P}$.\vs

The objective function of this optimization model is $\sum_{t=0}^{T-1} \hat{w}_t x_t$,
where $\hat{w}_t=(\sum_{d\in \overline{D}}w_t^d)/|\overline{D}|$ is the average PLF at time $t$
over a historical record of observations $\overline{D}$, eventually different from the training set.
Whenever forecasts are statistically reliable,
the objective function corresponds with the expected absolute PLF area of the band along the period $T$.
Defined so, the optimization is not instant-to-instant greedy,
in the sense that it could deteriorate the performance at some instants
in order to surpass the overall performance by gaining more in others,
which differentiates this work from related ones as \cite{WE:WE2129}.
In fact, this model doesn't need common hypothesis about the stochastic process,
such as Homogeneity or Markovianity.\vs

The second group of control variables is composed by those who determine which are the regular forecasts.
The variable $y_d \in \{0,1\}$ indicates whether the forecast issued on day $d$ should be considered regular ($y_d=1$),
or atypical ($y_d=0$). Unlike the $(x_t)$ variables, these new ones are binary.
We denote $D$ to the set of days for which the optimization problem is implemented. The complete model is as follows:

\begin{equation}\label{eq:minbandrel}
\left\{\begin{array}{l}
\min \displaystyle\sum_{t=0}^{T-1} \hat{w}_t x_t \vspace{0.1in}\\
\begin{array}{llc}
%z_t^d \ge w^d_t - p^d_t (1+x_t) - (1-y_d), & 0\le t\le T-1, d\in D, & (i)\vspace{0.05in}\\
%z_t^d \ge p^d_t (1-x_t) - w^d_t - (1-y_d), & 0\le t\le T-1, d\in D, & (ii)\vspace{0.05in}\\
p^d_t x_t - y_d + z_t^d \ge |w^d_t - p^d_t| - 1, & 0\le t\le T-1, d\in D, & (i)\vspace{0.05in}\\
\displaystyle\sum_{t=0}^{T-1} z_t^d \le T(\theta+1-y_d), & d\in D, & (ii)\vspace{0.02in}\\
\displaystyle\sum_{d\in D}y_d \ge \lambda D, & & (iii) \vspace{0.05in}\\
y_d\in\{0,1\}, \, 0 \leq x_t, z_t^d \leq 1.
\end{array}
\end{array}\right.
\end{equation}

\noindent The auxiliary variables $(z_t^d)$ account by how much the process ($w_t^d$) violates the confidence band around the forecast ($p_t^d$),
either at the top or the bottom, for those days classified as regular (i.e. when $y_d=1$).
For instance, if $y_d=1$ and $w_t^d \ge p_t^d (1+x_t)$,
then $z_t^d\ge w_t^d - p_t^d (1+x_t)\ge 0$ to satisfy equation $(i)$ in \eqref{eq:minbandrel} for that day $d$ at time $t$.
When $y_d=1$ and $w_t^d \le p_t^d (1-x_t)$, then $z_t^d$ should verify $z_t^d\ge p_t^d (1-x_t) - w_t^d\ge 0$ to satisfy equation $(i)$.
The optimization process pushes down the $z_t^d$ values,
which ultimately are to be set to $\max \left[ 0,w_t^d - p_t^d (1+x_t),p_t^d (1-x_t) - w_t^d\right]$.
That equation is always satisfied when $y_d=0$ by choosing $z_t^d=0$ for every $t$,
so atypical days are disregarded for violations.\vs

Given any day $d$, when $y_d=1$ (an effective day of the training set),
the second equation guarantees that the time-normalized cumulated off-band energy along the forecasted period $T$ is below $\theta$.
That is, in terms of the reliability: $1 - {\mathcal R}^{d} = (\sum_{t=0}^{T-1} z_t^d)/T \le\theta$;
so $\theta$ bounds the energy that lies outside the band to a fraction of the installed power plant.
As it happened with $(i)$, equation $(ii)$ is automatically satisfied when $y_d=0$.
Finally, equation $(iii)$ forces the problem to select at least $\lambda D$ days to be regular,
which combined with the persistence hypothesis conveys likelihood to the result.

\subsection{Experimental Evaluation}\label{sec:expresults}
The experimental evaluation of this work is based on the Uruguayan Electricity Market.
It uses public domain information available on the site: \url{http://www.ute.com.uy/SgePublico/ConsPrevGeneracioEolica.aspx}.
From this site, we have chosen two independent forecast sources: Garrad Hassan and Meteol\'ogica.
The third source (UTE pron\'ostico 3) was discarded because, by the time those forecasts were processed (early 2017),
the number of samples available was not large enough.
The data were pre-processed using a power assimilation methodology,
which fits forecasts along the first 6 hours in order to match the starting state ($w_0^d$).
The exact process is described in \cite{POWASS}.\vs

The used forecasts from Garrad Hassan were those issued at 1AM between April 5$^\mathrm{th}$, 2016, and March 10$^\mathrm{th}$, 2017.
All of those forecasts have a 72 hours horizon. Within this sample there are 302 days where both,
forecast and actual data are complete,
i.e., there are no missing records in any of those series.
Regarding the other provider (Meteol\'ogica), the number of complete records is 394,
with dates of issued ranging from January 1$^\mathrm{st}$, 2016, to March 10$^\mathrm{th}$, 2017.\vs

\begin{figure}[htpb!]
\centering
\includegraphics[width=0.75\paperwidth]{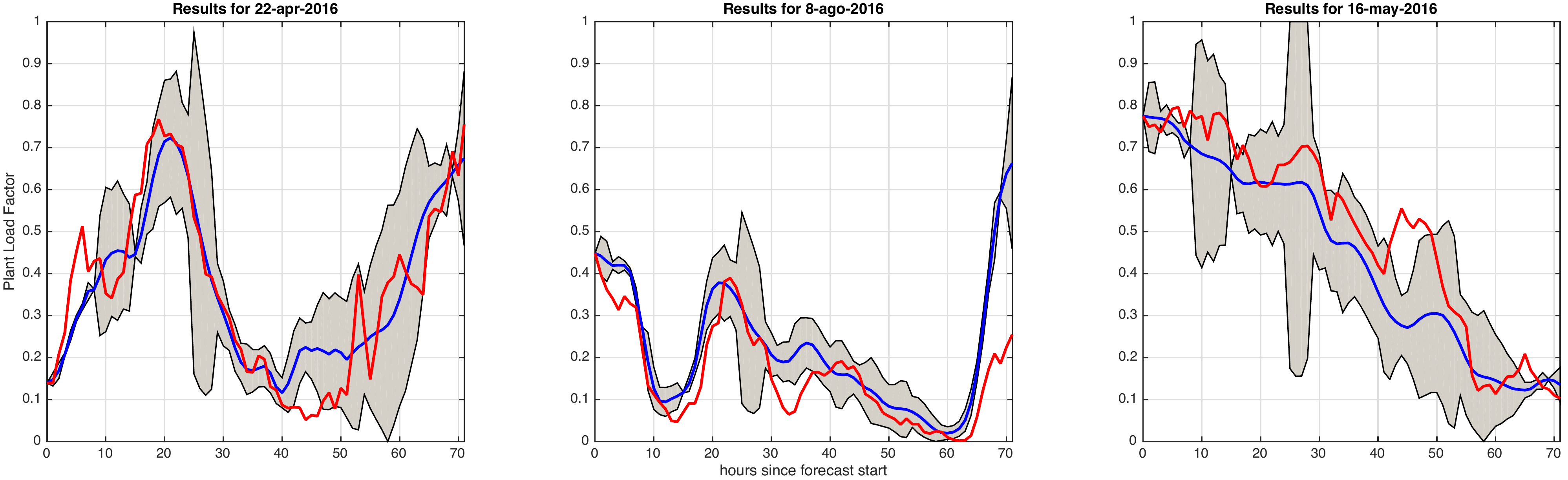}
\caption{Confidence bands for three days randomly choosed within Meteol\'ogica's training set [$\lambda$=1,$\theta$=0.05]}\label{fig:asingclust}
\vspace{-0.15in}
\end{figure}

Throughout this work, we used \verb#IBM(R) ILOG(R) CPLEX(R) Interactive Optimizer 12.6.3# as the optimizer.
The server was an \verb#HP ProLiant DL385 G7#, with 24 \verb#AMD Opteron(tm)# Processor 6172 with 64GB of RAM.
After running model \eqref{eq:minbandrel} over a training set comprising around 30\% of Meteol\'ogica's days (40\% of Garrad Hassan's)
we find bands like those sketched in \figref{fig:asingclust}.
The x-axis represents the number of hours ahead for each forecast,
while the y-axis corresponds to the PLF.
Blue curves are associated with forecasts while red ones are the actual values.
Finally, the grey area represents the confidence band for $\theta=0.05$ and $\lambda=1$.
The training was performed over a set $D$ of 120 randomly selected days out of a set of 394,
with the constraint that they belong to the historical registers of both forecasts providers.

\begin{figure}[htpb!]
\centering
\includegraphics[width=0.75\paperwidth]{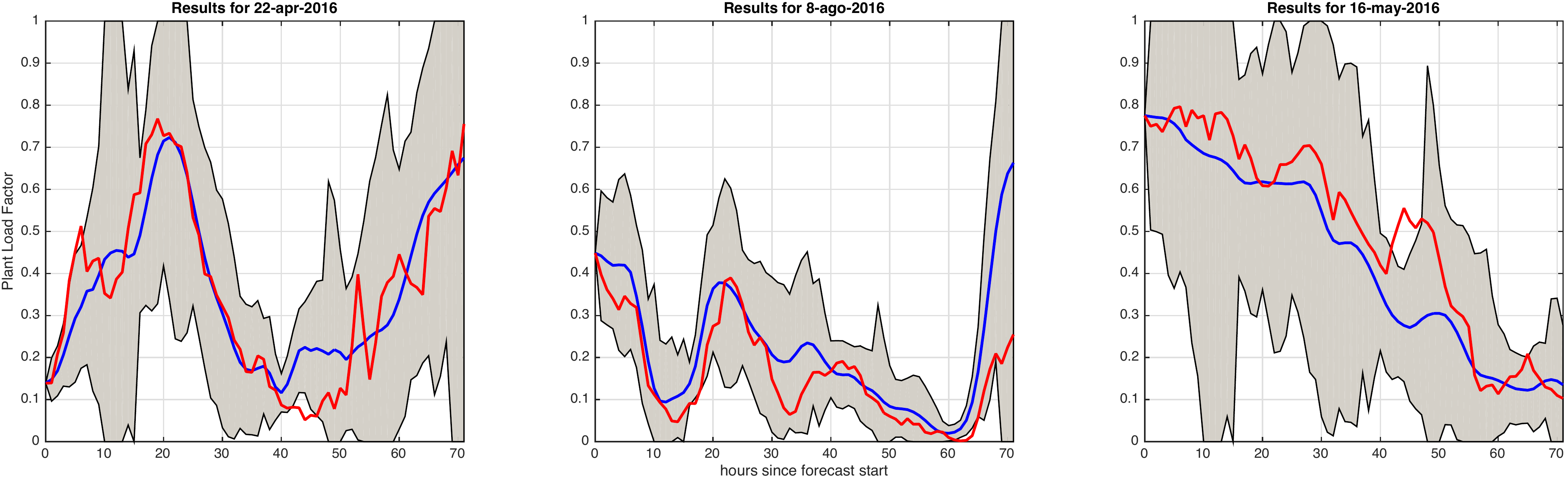}
\caption{Confidence bands for the same three days of \figref{fig:asingclust} when $\theta=0.01$ instead of $\theta=0.05$ [$\lambda$=1]}\label{fig:asingclust2}
\vspace{-0.15in}\end{figure}

\noindent Since $\lambda$ equals 1, every day within the training set must be effectively included;
that is, $y_d=1$ for each $d\in D$, so all days are treated as regular.
Furthermore, when $\lambda=1$ then \eqref{eq:minbandrel} turns out to be a pure linear programing problem,
and running times are within the second.
Fixed $\lambda$, it is of interest to explore how $\theta$ modifies the bands.
\figref{fig:asingclust2} shows the result over the same training set when $\theta=0.01$.

\noindent Observe that bands in \figref{fig:asingclust2} are wider than in \figref{fig:asingclust},
which is expected since we are less tolerant respect to how much energy (red curve) lies outside the band.
In order to balance reliability and thickness,
it is of interest to compute how much area do bands cover as we change $\theta$ while keeping $\lambda=1$.

\begin{figure}[htpb!]
\centering
\includegraphics[width=0.35\paperwidth]{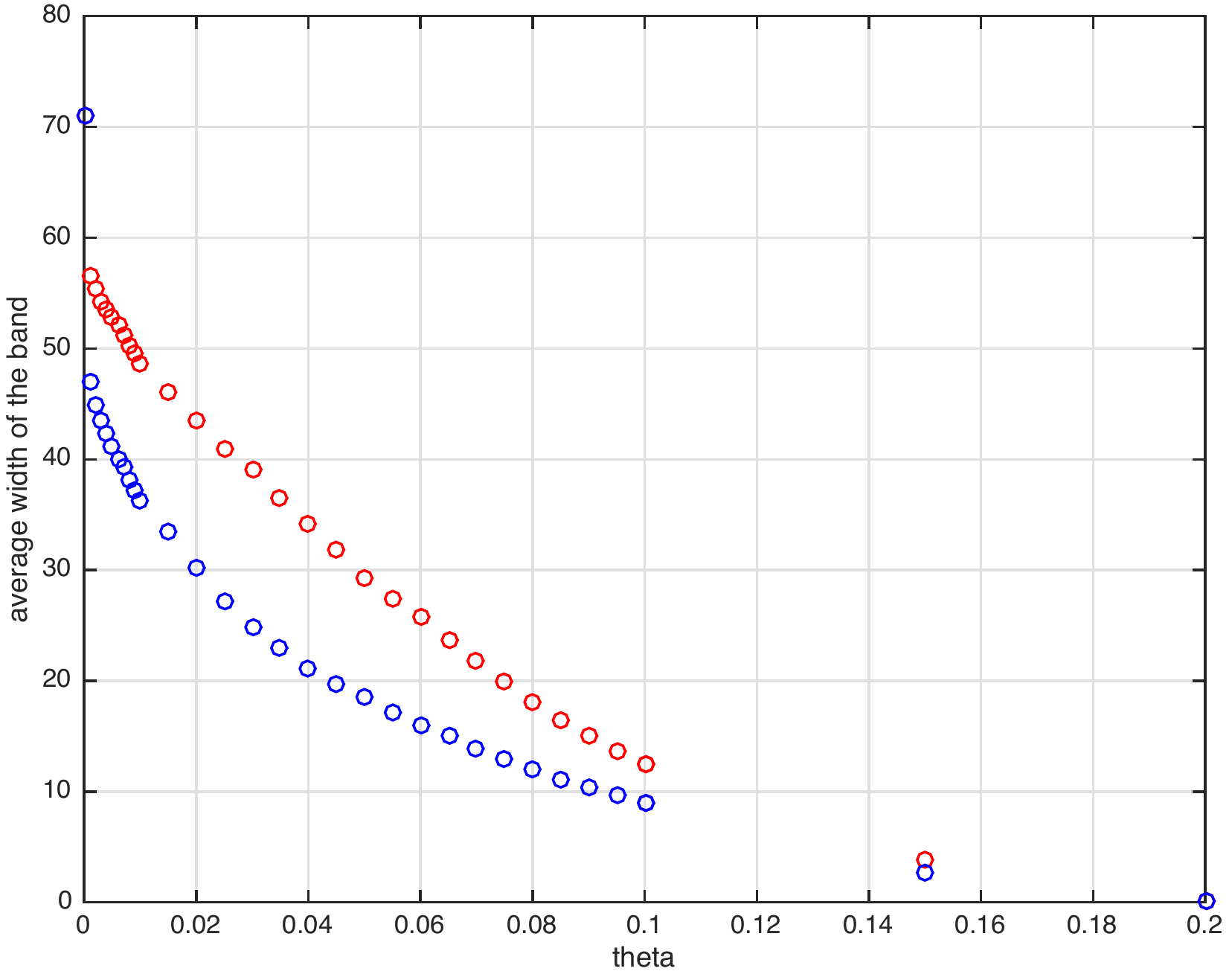}
\caption{Evolution of the average width of the band as $\theta$ changes while $\lambda$ is fixed in 1 [red samples are from Garrad Hassan
and blue ones are from Meteol\'ogica]}\label{fig:pareto}
\end{figure}

Experimental evaluation (see \figref{fig:pareto}) verifies that the average width of the confidence band,
when computed over the training set of forecasts,
falls down rapidly to 0, which is reached upon both companies when $\theta$ is close to 0.2.
Conversely, when $\theta\rightarrow 0$ the band tends to occupy the entire feasible set of trajectories,
an area of 71 which excludes $t=0$ since process and forecast always match at that point.
For any intermediate value, the quality of Meteol\'ogica's bands are always better than Garrad Hassan's.
Another parameter to consider is $\lambda$, which attends to the fact that whatever accurate a family of forecasts may be,
there will always be samples that degrades the overall quality of the whole.
\tabref{tab:resultsvslambda1} shows how some attributes of the bands change as $\lambda$ decreases from 1 to 0.8,
while $\theta$ remains fixed in 0.01.
The first four columns correspond to Meteol\'ogica forecasts and the second half does to Garrad Hassan's.
These metrics were computed outside the training set,
over a test set of 182 days common to both historical records.
Columns labeled as \emph{\% atypical} indicate the percentage of the samples, in each case,
whose off-band energies surpasses the 0.01 of the plant factor.
The columns $\overline{BW}$ and \%$\overline{BW}$ respectively show the average absolute and relative areas of the confidence bands over the test set,
using 72 as the full plant factor for the time horizon.
Finally, the number of seconds spent by the solver to find the optimal solution for each case
appears in the column labeled as \emph{t(secs)}.

\begin{table}[htpb!]
\caption{Experimentally estimated attributes for confidence $\theta=0.01$ bands as $\lambda$ decreases}\label{tab:resultsvslambda1}
\begin{threeparttable}
\begin{tabular}{c|cccr|cccr|}
\headrow
\thead{$\;\;\mathbf{\lambda}$} & \thead{\% atypical} & \thead{$\mathbf{\overline{BW}}$} & \thead{\%$\mathbf{\overline{BW}}$} & \thead{t (sec)} & \thead{\% atypical} & \thead{$\mathbf{\overline{BW}}$} & \thead{\%$\mathbf{\overline{BW}}$} & \thead{t (sec)} \\
\headcell{1.00} &   7.6\% & 34.4 & 47.8\% &     <1 &   9.2\% & 47.1 & 65.4\% &   <1 \\
\headcell{0.95} & 16.3\% & 29.8 & 41.1\% &   190 & 20.4\% & 41.5 & 57.6\% &  130 \\
\headcell{0.90} & 21.4\% & 26.2 & 36.4\% &   184 & 22.5\% & 40.1 & 55.7\% &  801 \\
\headcell{0.85} & 29.1\% & 24.1 & 33.5\% & 1312 & 25.0\% & 38.8 & 53.9\% & 3523 \\
\headcell{0.80} & 37.2\% & 22.3 & 31.0\% & 2188 & 29.6\% & 36.8 & 51.1\% & 5537 \\
\hline
\cellcolor{white} & \multicolumn{4}{c|}{\bf Meteol\'ogica} & \multicolumn{4}{c|}{\bf Garrad Hassan} \\
\cline{2-9}
\end{tabular}
%\begin{tablenotes}
%\item JKL, just keep laughing; MN, merry noise.
%\end{tablenotes}
\end{threeparttable}
\end{table}

\begin{figure}[htbp!]
\centering
\includegraphics[width=0.8\paperwidth]{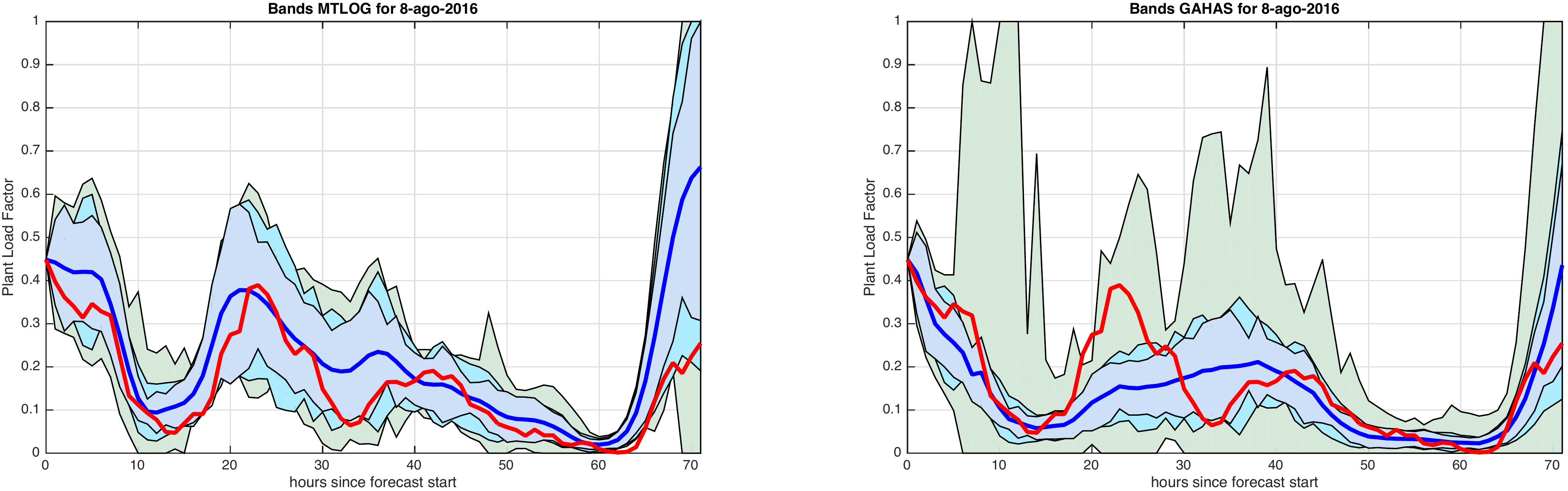}
\caption{Example bands for Meteol\'ogica (leftmost) and Garrad Hassan (rightmost),
computed for $\theta=0.01$ with $\lambda$ equal to: 1, 0.9 and 0.8, on August 8$^\mathrm{th}$,
2016 [red curves are actual power and blue are forecasts].}\label{fig:bandsexampl}
\vspace{-0.1in}
\end{figure}

\noindent To complement the table,
\figref{fig:bandsexampl} sketches bands for both companies at the same day (August 8$^\mathrm{th}$, 2016).
As the descriptive statistical information suggests,
the figure illustrates how much narrower and reliable Meteol\'ogica's bands are,
especially for lower values of $\lambda$.

\noindent The goal of this work was providing stochastic short-term optimal power dispatch schedulers with accurate confidence bands,
in the context of the Uruguayan Electricity Market.
In Uruguay, the average PLF is around 35\%,
so it is desirable for the expected width of confidence bands to be quite below that level;
otherwise, we would have to expect deviations of the actual process within the confidence band larger than the expected power itself.
We take 25\% of the PLF as the maximum allowed for the average width of the confidence band.

\noindent On the other hand, to keep the expected off-band energy below 10\% of the average wind power,
we should fix $\theta\le 0.1 \times 0.35=0.035$.
Besides, by late 2017, 35\% of the average wind power in the Uruguayan grid was coming from wind farms,
so the previous limit implies that the expected off-band power wind would be around 3.5\% of the total system's power.
Finally, we are willing to accept up to a 10\% of expected violations to that energy limit,
so the fraction of atypical days should be below 10\%.
In summary, our target figures are:
$\theta\le 0.035$, \%$\overline{BW} \le 25$\% and \emph{\%atypical}$\le$10\%.

\begin{table}[htpb!]
\caption{Experimentally estimated attributes for confidence $\theta=0.035$ bands as $\lambda$ decreases}\label{tab:resultsvslambda3}
\begin{threeparttable}
\begin{tabular}{c|cccr|cccr|}
\headrow
\thead{$\;\;\mathbf{\lambda}$} & \thead{\% atypical} & \thead{$\mathbf{\overline{BW}}$} & \thead{\%$\mathbf{\overline{BW}}$} & \thead{t (sec)} & \thead{\% atypical} & \thead{$\mathbf{\overline{BW}}$} & \thead{\%$\mathbf{\overline{BW}}$} & \thead{t (sec)} \\
\headcell{1.00} &   6.6\% & 21.3 & 29.6\% &     <1 &   5.6\% & 34.3 & 47.6\% &   <1 \\
\headcell{0.95} & 14.8\% & 16.0 & 22.2\% &   117 & 12.2\% & 26.8 & 37.2\% &  129 \\
\headcell{0.90} & 20.4\% & 13.8 & 19.2\% &   138 & 16.3\% & 24.8 & 34.3\% &  428 \\
\headcell{0.85} & 27.1\% & 12.5 & 17.4\% &   265 & 22.5\% & 22.1 & 30.7\% &  272 \\
\headcell{0.80} & 32.1\% & 11.7 & 16.3\% &   664 & 25.5\% & 19.7 & 27.4\% & 1518 \\
\hline
\cellcolor{white} & \multicolumn{4}{c|}{\bf Meteol\'ogica} & \multicolumn{4}{c|}{\bf Garrad Hassan} \\
\cline{2-9}
\end{tabular}
%\begin{tablenotes}
%\item JKL, just keep laughing; MN, merry noise.
%\end{tablenotes}
\end{threeparttable}
\end{table}

\noindent Observe that no record in \tabref{tab:resultsvslambda1} or in \tabref{tab:resultsvslambda3} fulfills all of those constraints simultaneously.
The strategy to follow in order to overcome that problem consists in combining bands of adequate width, but with a higher than expected percentage of atypical days,
into a new one with improved confidence.

\section{COMBINING FORECASTS}\label{sec:combfore}

At first sight we might think that a convex combination of forecasts and their bands would inherit the width of each one,
and that we cannot improve band's quality by combining them.
However, since forecasts tend to compensate each other and bands are truncated to fit within [0,1],
this is not necessarily so, and we can in fact improve the expected width by combining bands.

\begin{figure}[htbp!]
\centering
\includegraphics[width=0.75\paperwidth]{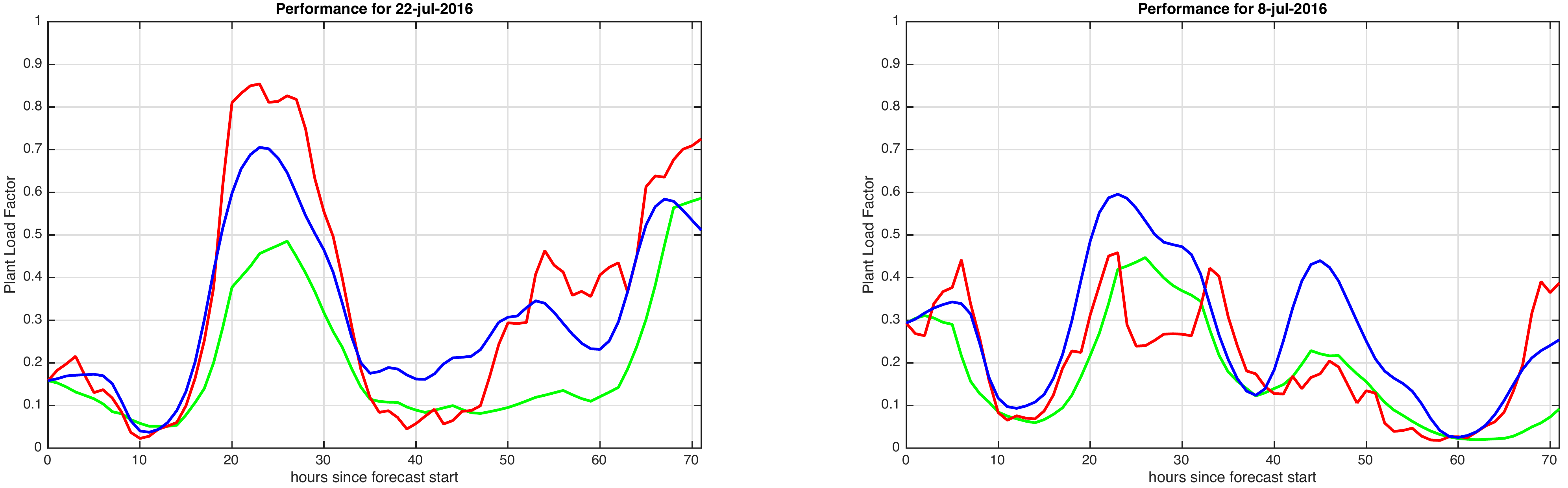}
\caption{Atypical forecasts of both companies,
regarding the distance to the actual power 
[Red curves correspond to actual process, blue ones to Meteol\'ogica and green ones to Garrad Hassan]}\label{fig:atypicalGM}
\end{figure}

\noindent Besides, we indeed could improve its confidence if atypical days were somehow independent,
since a combination of atypical situations in both bands would be rarer than any of them by separate.
In \figref{fig:atypicalGM} are presented two days specifically selected to show that effect:
Jule 22$^\mathrm{nd}$, 2016, and Jule 8$^\mathrm{th}$, 2016.
On the leftmost chart it can be observed that the distance between the actual process and Garrad Hassan's forecast (i.e. red and green curves) is large,
so any optimization forced to fit the process within the band should produce a too much wide one.
An analog situation occurs over the rightmost chart,
although in this case the forecast that is farthest from the process is that of Meteol\'ogica (blue curve).
By lowering $\lambda$ to allow the optimization process to discard atypical samples,
we can find narrower bands, since now they are less affected by samples that are intrinsically bad.
Later on, by combining independent bands we could regain the confidence lost after decreasing $\lambda$.
To check the consistency of this idea, we analyze how independent atypical days are.

\begin{table}[htpb!]
\caption{Percentages of atypical days and their correlations for different combinations of $\lambda$'s and $\theta$'s}\label{tab:independence}
\begin{threeparttable}
%\begin{tabular}{c|cccc|cccc|}
\begin{tabular}{c|crrrc|crrrc|}
\headrow
\thead{$\;\;\;\mathbf{\lambda}$} & \thead{$\;\;\;\mathbf{\theta}$} & \thead{Meteo.} & \thead{Ga. Hass.} & \thead{Simult.} & \thead{$\mathbf{\rho_{Mt,Gh}}$} &
                                                    \thead{$\;\;\;\;\mathbf{\theta}$} & \thead{Meteo.} & \thead{Ga. Hass.} & \thead{Simult.} & \thead{$\mathbf{\rho_{Mt,Gh}}$} \\
\headcell{1.00} & \headcell{0.01} &   7.6\%~~ &   9.2\%~~ &   3.1\%~~~ & 0.307 & \headcell{0.035} &   6.6\%~~ &   5.6\%~~ &   2.0\%~~~ & 0.291 \\
\headcell{0.95} & \headcell{0.01} & 16.3\%~~ & 20.4\%~~ &   7.1\%~~~ & 0.256 &\headcell{0.035} & 14.8\%~~ & 12.2\%~~ &   5.1\%~~~ & 0.282 \\
\headcell{0.90} & \headcell{0.01} & 21.4\%~~ & 22.5\%~~ &   9.2\%~~~ & 0.255 &\headcell{0.035} & 20.4\%~~ & 16.3\%~~ &   8.2\%~~~ & 0.324 \\
\headcell{0.85} & \headcell{0.01} & 29.1\%~~ & 25.0\%~~ & 12.8\%~~~ & 0.279 &\headcell{0.035} & 27.1\%~~ & 22.5\%~~ & 10.7\%~~~ & 0.251 \\
\headcell{0.80} & \headcell{0.01} & 37.2\%~~ & 29.6\%~~ & 15.8\%~~~ & 0.217 &\headcell{0.035} & 32.1\%~~ & 25.5\%~~ & 13.3\%~~~ & 0.249 \\
\hline
\end{tabular}
%\begin{tablenotes}
%\item JKL, just keep laughing; MN, merry noise.
%\end{tablenotes}
\end{threeparttable}
\end{table}

\noindent\tabref{tab:independence} recapitulates figures of atypical days under different combinations of $\lambda$ and $\theta$.
It also adds information regarding combinations of atypical occurrences,
as well as estimators for the correlation between Bernoulli's random variables $Mt(d)$ and $Gt(d)$,
which evaluates to 1 (resp. 0) if and only if the forecast for day $d$ of the respective company classifies as atypical (resp. regular).
From those correlation values, we infer that atypical days of both companies are positive but weakly correlated\footnote{By running simple simulations with two sets of independent Bernoulli's random variables with the same expected value, one can observe that the correlations values of the table appeared when 1 of between 3 and 4 samples of one set copied the value of the other.}. \vs

Given two forecasts $\hat{p^d}$ and $\check{p^d}$ and the respective coefficients ($\hat{x_t}$, $\check{x_t}$) to compute their confidence bands,
an easy method to craft new forecasts and bands is by a convex combination $p^d_t=\alpha\hat{p^d_t}+(1-\alpha)\check{p^d_t}$, where $0\le\alpha\le 1$.
Theoretically, we could explore combinations of $\theta$, $\lambda$ and $\alpha$ in order to maximize our expectations for the outcome.
However, such exploration is not that simple in terms of computations.
As soon as $\lambda$ decreases, problem \eqref{eq:minbandrel} becomes combinatorial,
and times necessary to find solutions increases substantially as it is accounted in columns $t(sec)$ of \tabref{tab:resultsvslambda1} and \tabref{tab:resultsvslambda3}.
Therefore, in this work we limit to manually explore a handful of alternatives,
just to illustrate about how the technique allows to improve the quality.
First of all, we set $\theta$ to 0.035, i.e., our target of \emph{regular confidence}.
In the second place, we estimate bounds to $\alpha$ values to satisfy the aimed average width of the band,
as a combination of their respective widths.
Checking on \tabref{tab:resultsvslambda3} we realize that $\lambda=1$ is not viable,
because widths of both companies (Meteol\'ogica and Garrad Hassan) are above the target (25\%), so it will be their combinations.
For $\lambda=0.95$ we can estimate an upper bound for the outcome's width by $0.222\alpha+0.372(1-\alpha)$,
and since $0.222\alpha+0.372(1-\alpha)\le 0.25$ we conclude that $\alpha\ge 0.813$,
which means that the outcome would be mostly (81\%) Meteol\'ogica's.
How this translates to results is not that linear.
After exploring over values of $\alpha$ for different $\lambda$'s,
we concluded that the values that minimize the expected width of the band over the training-set,
while keep the percentage of atypical days under the 10\% threshold,
are $\alpha=0.91$ for $\lambda=0.95$ and $\alpha=0.78$ for $\lambda=0.90$,
whose expected relative band's widths are: 23.9\% and 23.0\%, both below the 25\% target.
We couldn't find feasible combinations for lower values of $\lambda$.
Thus, we choose the second set as the optimal combined band.\vs

%%% Performance of optimal combined band %%%
\subsection{Performance of optimal combined band}

To analyze the performance of the combined band we present qualitative and quantitive evidence.

\begin{figure}[htpb!]
\centering
\includegraphics[width=0.75\paperwidth]{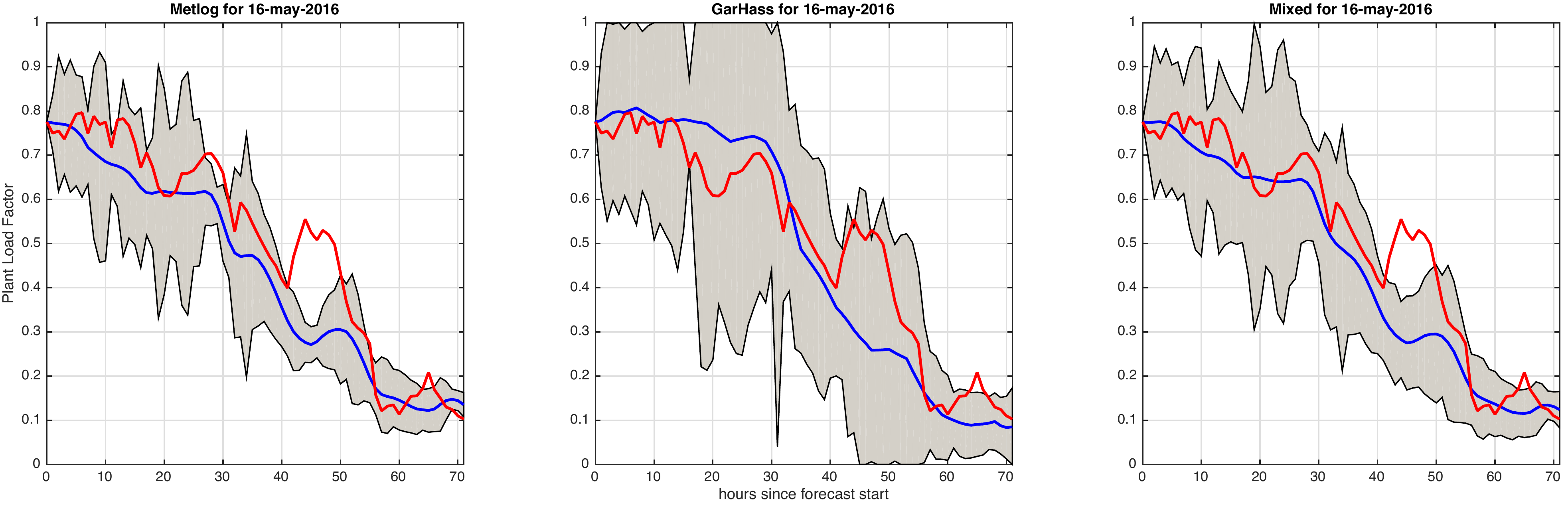}
\caption{Confidence bands for dissociated and combined forecast at 16$^\mathrm{th}$ May, 2016 [$\lambda$=0.90, $\theta$=0.035]}\label{fig:mx16may}
\vspace{-0.15in}
\end{figure}

\noindent\figref{fig:mx16may} shows bands as computed for Meteol\'ogica (leftmost), Garrad Hassam (middle) and the optimal combination,
when computed using $\lambda$=0.90 and $\theta$=0.035, in a random day.
It can be checked that the third improves the confidence regarding the first,
with a notoriously better band than the second.

\begin{figure}[htpb!]
\centering
\includegraphics[width=0.75\paperwidth]{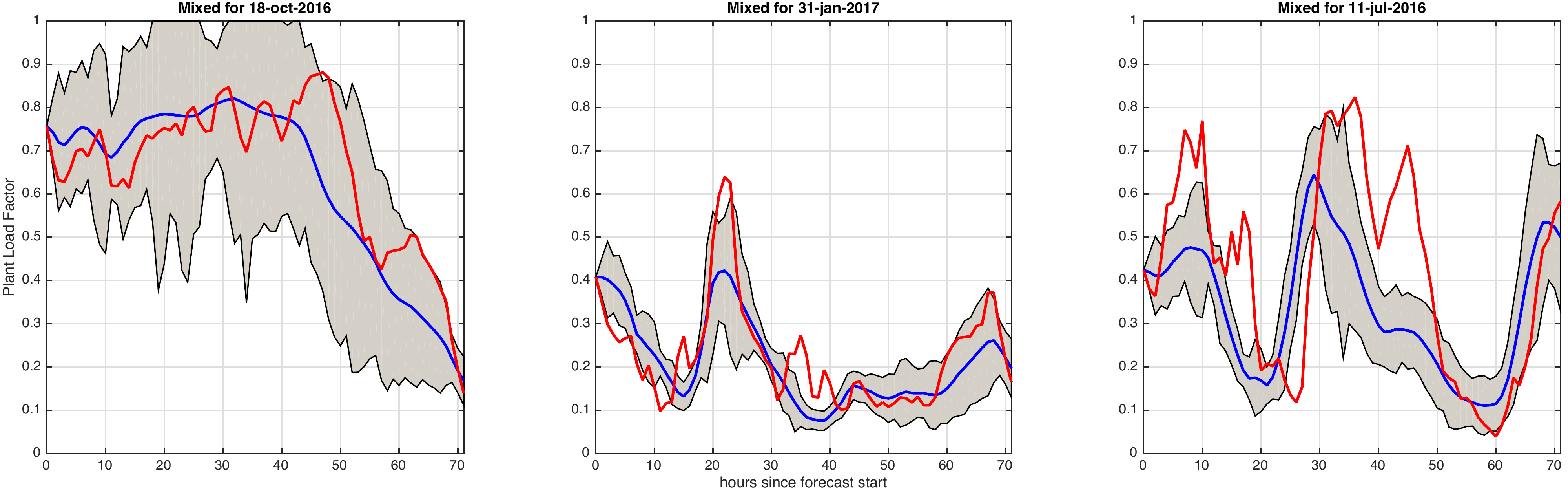}
\caption{Confidence bands for some extreme cases [$\lambda$=0.90, $\theta$=0.035 and $\alpha$=0.78]}\label{fig:mxbdmdwd}
\vspace{-0.15in}
\end{figure}

\noindent All the bands in \figref{fig:mxbdmdwd} correspond to the optimal combination of both forecast ($\alpha=0.78$, $\theta$=0.035 and $\lambda=0.90$).
The leftmost chart, corresponding to 18$^\mathrm{th}$ October, 2016, examples a very reliable case although it has a band of regular quality.
The middle chart, elaborated upon forecasts of 31$^\mathrm{st}$ January, 2017, shows a sample with average reliability and a much better band,
while the rightmost (11$^\mathrm{th}$ July, 2016) corresponds to an atypical day, in fact,
to that with the worst disagreement in the off-band energy within the test-set.\vs

\begin{figure}[htpb!]
\centering
\includegraphics[width=0.8\paperwidth]{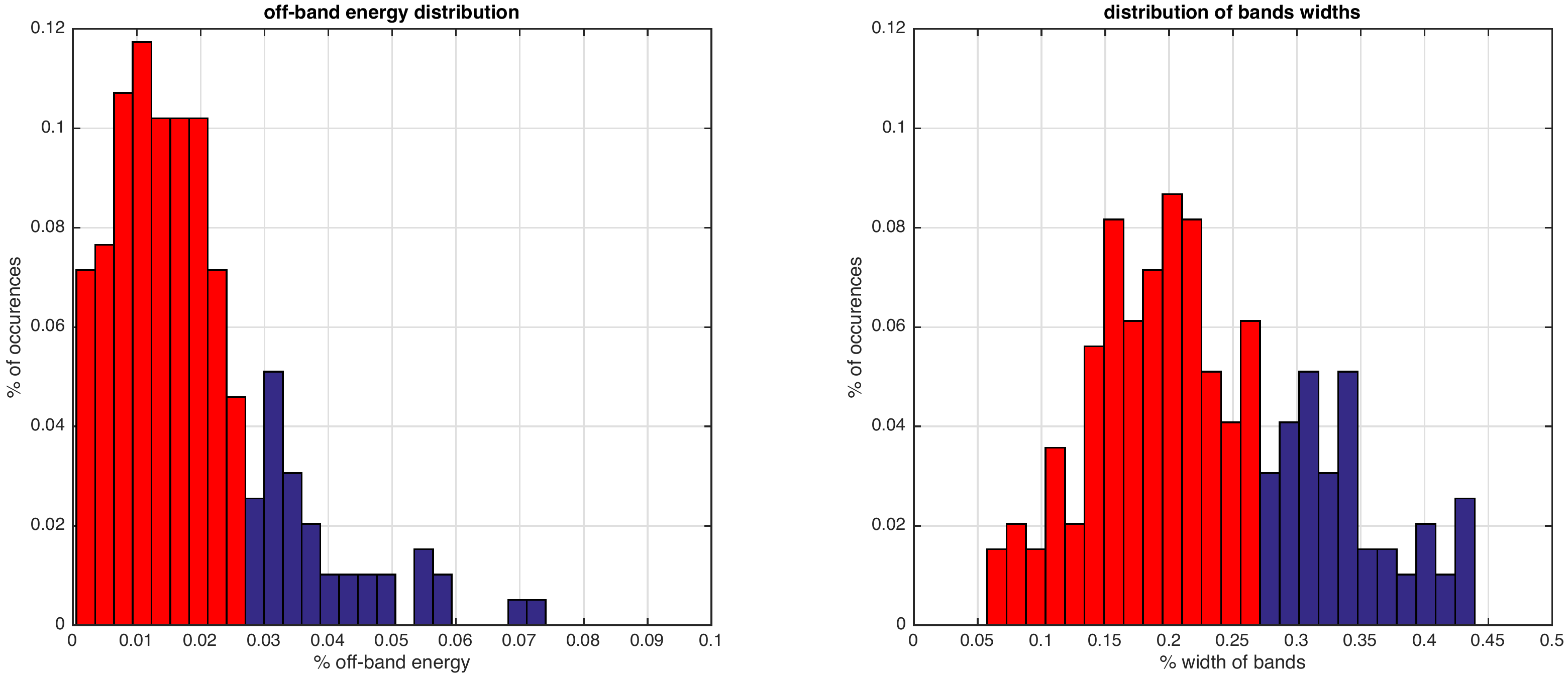}
\caption{Histograms for relative off-band energy and widths of the bands}\label{fig:histodev}
\vspace{-0.15in}
\end{figure}

It is worth wandering how much energy lies outside the confidence band when violations happen,
and how narrow confidence bands are.
\figref{fig:histodev} shows histograms computed up from the test set.
The leftmost corresponds to the distribution of the off-band energy normalized by the total PLF along the period (72).
It is observed that no sample disagrees in more than 7.3\%,
while in 75\% of the samples (those colored with red) that percentage is lower than 2.6\%.
The combined disagreements of samples associated with red bars add up to 50\% of the total off-band energy of the test set.
These results reflect the quality of the forecasts and bands computed by the method.

\noindent The rightmost histogram of \figref{fig:histodev} presents the distribution of widths for the bands in the test set,
which are a measure of the quality.
We find out that the width for some bands gets close to 45\%,
which are bad results. Such situations however, seldom happen.
In fact 66\% of the samples in the test set (those remarked with red bars)
have relative widths lower than 26.4\%,
and all of those combined represent the half of the total width over the set.\vs

We conclude then that the result is not only satisfactory regarding our initial average performance goals
($\theta\le 0.035$, \%BW$\le$25\% and \%atypical$\le$10\%),
but it is pretty good in terms of the overall quality of the bands
and specially in terms of the energy confidence of them.

%\begin{table}[htpb!]
%\caption{Experimentally estimated attributes for confidence $\theta=0.05$ bands as $\lambda$ decreases}\label{tab:resultsvslambda2}
%\begin{threeparttable}
%\begin{tabular}{c|cccr|cccr|}
%\headrow
%\thead{$\;\;\mathbf{\lambda}$} & \thead{\% atypical} & \thead{$\mathbf{\overline{BW}}$} & \thead{\%$\mathbf{\overline{BW}}$} & \thead{t (sec)} & \thead{\% atypical} & \thead{$\mathbf{\overline{BW}}$} & \thead{\%$\mathbf{\overline{BW}}$} & \thead{t (sec)} \\
%\headcell{1.00} &    4.1\% & 17.0 & 23.6\% &  <1 &   5.6\% & 27.3 & 37.9\% &   <1 \\
%\headcell{0.95} &  11.7\% & 12.0 & 16.7\% &   76 & 12.2\% & 20.1 & 27.9\% &  111 \\
%\headcell{0.90} &  17.4\% &  9.7 & 13.5\% &  148 & 14.8\% & 19.0 & 26.4\% &  209 \\
%\headcell{0.85} &  24.0\% &  8.6 & 11.9\% &  202 & 20.4\% & 16.3 & 22.6\% &  294 \\
%\headcell{0.80} &  31.1\% &  7.8 & 10.8\% &  532 & 26.5\% & 14.1 & 19.6\% &  371 \\
%\hline
%\cellcolor{white} & \multicolumn{4}{c|}{\bf Meteol\'ogica} & \multicolumn{4}{c|}{\bf Garrad Hassan} \\
%\cline{2-9}
%\end{tabular}
%%\begin{tablenotes}
%%\item JKL, just keep laughing; MN, merry noise.
%%\end{tablenotes}
%\end{threeparttable}
%\end{table}

%%%%%%%%%%%%%%%%%%%%%%%%%%
%%% CONCLUSIONS AND FUTURE WORK %%%
%%%%%%%%%%%%%%%%%%%%%%%%%%
%%%%%%%%%%%%%%%%%%%%%%%%%%
%%% CONCLUSIONS AND FUTURE WORK %%%
%%%%%%%%%%%%%%%%%%%%%%%%%%
\section{CONCLUSIONS AND FUTURE WORK}\label{sec:concl}

In this work, we introduce a novel optimization technique for building confidence bands around forecasts of wind power generation over a horizon of 72 hours. The analysis is based on the historical data set provided by the Uruguayan Electricity Market. The technique allows to discard a portion of atypical days in the training set, while controls the average cumulated energy that lies outside bands. With an appropriate choice of the parameters involved in the analysis, the model is successful in providing bands satisfying natural requirements on confidence and width. \vs

A remarkable conclusion of this work is that the use of an optimal convex combination of independent forecasts and its corresponding confidence bands improves significantly the performance of the model. For instance, the experimental evaluation of \secref{sec:bandopt} 
suggests that Meteol\'ogica forecasts performs, in average, better than Garrad Hassan's. However, an appropriate convex combination of both provides better results: even when most of the weight of the combination goes to Meteol\'ogica, the inclusion of Garrad Hassan forecast conveys stability to the result, compensating the fact that some atypical days in the forecast provided for Meteol\'ogica are regular according to Garrad Hassan. This idea could be extended of course including more forecasts providers. \vs

A drawback of the analysis that we mention here is that available data set at the moment of this work was not too large (around 300 days). We expect that the performance of the technique will work even better with a complete data set of a few years. However, this size introduces a problem: increasing the training set significantly increases computation times. So, in such situation is necessary the introduction of heuristic algorithms to solve the optimization problem. Even more if the parameters of convex combination of forecasts are also included in the analysis. A previous clusterization of forecasts might also improve the performance.\vs

%Finally, we think it is worth trying to mix this technique with parametric stochastic differential equations models,
%modifying the diffusive term associated with the Wiener process in order to keep its trajectories within confidence bands as presented in this paper.
%In \cite{EKTempone} and \cite{FOR:FOR2367}, the process is forced to exist within [0,1] (the PLF),
%but models could be modified to keep the process between functions $lb(t)$ and $ub(t)$ (i.e. band's limits),
%obtained by smoothly interpolating our discrete set of point, for instance using splines.
%Such approach would make possible to use continuous time optimization models to trustworthily solve the short-term dispatch of the grid.
%This is part of future work that we pretend to study in the next months.\vs

Finally, we think it is worth trying to mix this technique with parametric stochastic differential equations models,
modifying the diffusive term associated with the Wiener process in order to keep its trajectories within confidence bands as those presented in this paper.
In \cite{EKTempone} and \cite{FOR:FOR2367}, the process is forced to exist within [0,1] (the PLF),
but models could be modified to keep the process between functions $lb(t)$ and $ub(t)$ (i.e. limits of the bands),
obtained by smoothly interpolating our discrete set of point, for instance using splines.
Such approach would make possible to use continuous time optimization models to trustworthily solve the short-term dispatch of the grid.
This is part of future work that we pretend to study in the next months.

\section*{acknowledgements}
This work was partially supported by PEDECIBA-Inform\'atica (Uruguay), by the STIC-AMSUD project 15STIC-07 DAT (joint project Chile-France-Uruguay),
and by ANII (Agencia Nacional de Investigaci\'on e Innovaci\'on, Uruguay).
The authors want to acknowledge to: Gerardo Rubino (senior researcher at Inria’s center in Rennes, were most of this paper was written), Ra\'ul Tempone, H\'ector Cancela, Marco Scavino, Franco Robledo, Alfredo Piria and Juli\'an Viera, for fruitful discussions during the realization of this work.

%Work partially supported by PEDECIBA-Inform\'atica (Uruguay), by the STIC-AMSUD project 15STIC-07 DAT (joint project Chile-France-Uruguay),
%and by ANII (Agencia Nacional de Investigaci\'on e Innovaci\'on, Uruguay). The authors want to acknowledge Gerardo Rubino (senior researcher at Inria’s center in Rennes, were most of this paper was written), H\'ector Cancela, Ra\'ul Tempone, Marco Scavino, Alfredo Piria and Juli\'an Viera,
%for fruitful discussions during the realization of this work.

%\section*{conflict of interest}
%You may be asked to provide a conflict of interest statement during the submission process. Please check the journal's author guidelines for details on what to include in this section. Please ensure you liaise with all co-authors to confirm agreement with the final statement.

%\printendnotes

% Submissions are not required to reflect the precise reference formatting of the journal (use of italics, bold etc.), however it is important that all key elements of each reference are included.
%\bibliography{sample}
\bibliography{Risso_Guerberoff_2018_1_arXiv}

%\newpage

\begin{biography}[risso]{Claudio Risso}
holds degrees in Informatics and Electrical Engineering,
an MSc in Mathematical Eng. and a PhD in Computer Science,
Claudio Risso's academic research focuses on Combinatorial Optimization.
While most of his research was conducted on network design problems,
Dr. Risso is recently exploring how to apply exact and heuristic optimization methods in other areas,
such as data analysis and statistics. This is his first job in renewable energy forecasting.
%\bigskip
%\bigskip
\end{biography}

\begin{biography}[guerberoff]{Gustavo Guerberoff}
received his Ph. D. in Physics from Universidad Nacional de C\'ordoba (Argentina) in 1997.
He was a postdoc at Universidade de S\~ao Paulo (Brazil) and Universidad de Chile (Chile).
%From 2002 he has a professor position at Facultad de Ingenier\'ia, Universidad de la Rep\'ublica (Uruguay). 
His research interests cover different areas: Statistical Mechanics, Probability,
and Stochastic Processes applied to Biology, Bioinformatics and Engineering.
This is his first job in renewable energy forecasting. 
%\bigskip
%\bigskip
\end{biography}

%\newpage
\graphicalabstract{juanpabloterra}{The increasing rate of penetration of non-conventional renewable energies
is affecting the traditional assumption of controllability over energy sources.
Power dispatching scheduling methods need to integrate the intrinsic randomness of some new sources,
among which, wind energy is particularly difficult to treat.
This work aims on the construction of confidence bands around wind energy forecasts.
Complementarily, the proposed technique can be extended to integrate multiple forecasts into a single one,
whose band's width is narrower at the same level of confidence.
The work is based upon a real-world application case,
developed for the Uruguayan Electricity Market,
a world leader country in the penetration of renewable energies.
%Please check the journal's author guildines for whether a graphical abstract, key points, new findings, or other items are required for display in the Table of Contents.
}

\end{document}